\documentclass[twocolumn,amssymb,fleqn]{revtex4} %fleqn: make all eqns left alighed  ,showpacs
\usepackage{epsfig,amssymb,amsmath,graphicx,hyperref}  
\usepackage{color}

\usepackage{multirow} 
\usepackage{tabularx}

\usepackage{graphicx}
\graphicspath{{img/}}
\usepackage[caption=false]{subfig}
\newcommand{\be} {\begin{eqnarray}}
\newcommand{\ee} {\end{eqnarray} }

\newcommand{\f} {\frac }

\newcommand{\lb} {\left( }
\newcommand{\rb} {\right) }

\begin{document}

\title{Plastic instability of annular crystalline membrane in circular confinement}
\author{Honghui Sun and Zhenwei Yao}
\email{zyao@sjtu.edu.cn}
\affiliation{School of Physics and Astronomy, and Institute of Natural
Sciences, Shanghai Jiao Tong University, Shanghai 200240, China}
\begin{abstract} 
Understanding the mechanical instabilities of two-dimensional membranes has strong
  connection to the subjects of structure instabilities, morphology control and
  materials failures. In this work, we investigate the plastic mechanism
  developed in the annular crystalline membrane system for adapting to the
  shrinking space, which is caused by the controllable gradual expansion of the
  inner boundary. In the process of plastic deformation, we find the continuous
  generation of dislocations at the inner boundary, and their collective migration to the
  outer boundary; this neat dynamic scenario of dislocation current captures the
  complicated reorganization process of the particles. We also reveal the
  characteristic vortex structure arising from the interplay of topological
  defects and the displacement field. These results may find applications in the
  precise control of structural instabilities in packings of particulate
  matter and covalently bonded systems.
\end{abstract}

\maketitle

\section{Introduction}

Mechanical instability of two-dimensional membranes in confined geometry is a
common phenomenon in nature and
industry~\cite{Instabilities2008,audoly2010elasticity,bostwick2016elastic}. The
adaptation of the membrane to the confined space leads to a wealth of
morphologies across length scales ranging from mesoscopic (e.g., fluid
membranes and polymerized membranes) to macroscopic (e.g., paper,
metal foils, tree leaves, and flower petals) that are closely related to
biological applications and materials
design~\cite{cerda2003geometry,Marder2007,li2012mechanics,PhysRevLett.111.174302,bostwick2016elastic}.
Depending on the constituents composing the membrane and their interaction, the
mechanical instability could be classified into elastic and plastic
categories.  Much has been learned about the nature of elastic instability by
mechanical and statistical analysis of the emergent structures arising in
elastic
membranes~\cite{kantor1986statistical,lobkovsky1995scaling,Nelson2004c,chen2010minimal,balankin2010fractal,Davidovitch2011,king2012elastic,Grason2013}.
An important illustrative example is the crumpling of a thin sheet within a
confining container of specific shape or by
hand~\cite{gomes1989mechanically,balankin2010fractal,fokker2019crumpling}. A
series of characteristic structures are developed in the self-adaptation process
of the sheet, including developable cones (or
d-cones)~\cite{ben1997crumpled,cerda1998conical,mora2002thin,cerda2005confined},
ridges~\cite{lobkovsky1995scaling,lobkovsky1997properties,matan2002crumpling},
and
folds~\cite{matan2002crumpling,vliegenthart2006forced,tallinen2008deterministic}.
Crumpled structures may serve as a basis for creating robust mechanical
metamaterials for their desirable mechanical
properties~\cite{fokker2019crumpling,gimenez2023crumpled}.

Plastic instability represents a distinct mechanism for restructuring the
membrane~\cite{cottrell1965dislocations,fleck1994strain,nicola2006plastic,Grason2013,negri2015deformation,chen2022geometry}.
In contrast to elastic instability, the plasticity phenomenon occurring in the
regime of large deformation has not been explored thoroughly. Understanding the
plastic instability, which cannot be avoided in real
systems~\cite{weil1955large,matan2002crumpling,chopin2016disclinations}, has
strong connection to the subjects of structure instabilities and materials
failures~\cite{chopin2016disclinations,celli2020compliant}.  The confluence of
experimental and theoretical investigations on the model of the curved 2D
crystal fabricated by packings of particles shows the profound role of
topological defects in the disruption of crystalline order, and suggests that
topological defects offer a unique perspective for understanding
plasticity~\cite{Nelson1987,nelson2002defects,bowick2009two,wales2014chemistry}.

The goal of this work is to explore the plastic instability in the annular
crystalline membrane system. The model consists of Lennard-Jones (L-J) particles
in triangular lattice confined in an annulus. The schematic plot of the annular
membrane model is shown in Fig.~\ref{sch}. The featured energy minimum structure
in the L-J potential curve enables the breaking of the bonds and thus allows one
to capture the plastic deformation process by analyzing the underlying
topological defect structure~\cite{jones1924determination}. The deformation of
the annular membrane is driven by the controllable gradual expansion of the
inner boundary; the outer boundary is anchored. Note that the elastic
deformations of the annular membrane by shrinking or twisting the inner boundary
have been
investigated~\cite{geminard2004wrinkle,plaut2009linearly,zhang2012tunable,huang2019fourier}.
Upon the expansion, the annular geometry of the membrane naturally leads to
focused stress near the inner boundary, which could trigger mechanical
instabilities of different kinds depending on the interaction of the particles
composing the membrane. In connection to applications, the subject of the
plasticity instability of the annular membrane system is related to a series of
biomechanical and mechanical engineering problems, such as the healing of skin
wounds~\cite{cerda2005mechanics}, and the mechanical effect of circular holes
introduced in clinical procedures~\cite{david2004redistribution}.

The main results of this work are presented below. Based on the L-J lattice
model, we first resort to the combination of analytical elasticity theory and
numerical simulations to explore the elastic regime of small deformation and to
confirm the reliability of the computational approach, which is used to
explore the regime of plastic deformation. In the large deformation regime of
interest, the reorganization of the particles is analyzed from the
perspective of topological defect. Upon the gradual expansion of the inner
boundary, we observe the collective migration of the continuously generated
dislocations at the inner boundary to the outer boundary. The reorganization
process of the particles is captured by the dynamic scenario of the dislocation
current. We also identify the irregular spots in the displacement field that
allow one to pinpoint the locations of the ensuing topological defects, and
reveal the characteristic vortex structure arising from the interplay of
topological defects and the displacement field. These results may have potential
practical consequences in the precise engineering of structural instabilities in
packings of particulate matter and covalently bonded systems.

\section{Model and method}

\begin{figure}[t]  % h: put figure just in this position if possible
\centering 
\includegraphics[width=2in]{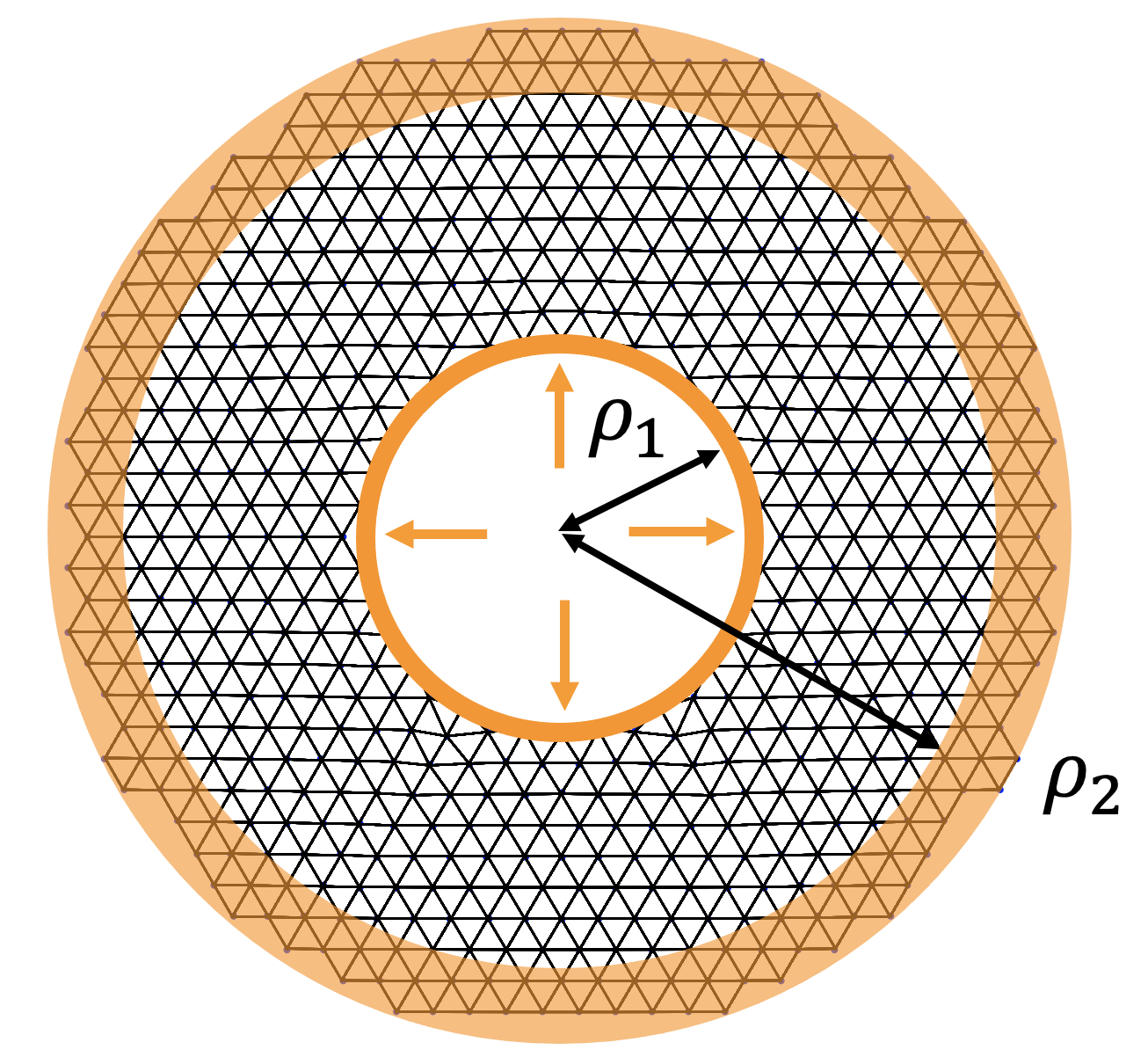}
  \caption{Schematic plot of the 2D annular crystalline membrane. The model
  consists of Lennard-Jones particles in triangular lattice confined in the
  annulus. The 2D deformation of the annular membrane is driven by the gradual
  expansion of the inner boundary, whose radius is denoted as $\rho_1$ and
  $\rho'_1$ in the initial stress-free state and the deformed state,
  respectively. To avoid the introduction of topological defects at the outer
  boundary, the anchored particles within the outer annulus (in orange) are in
  triangular lattice that is commensurate with the lattice of the annular
  membrane.  }
\label{sch}
\end{figure}

We resort to the L-J lattice models to explore the 2D plastic deformation of
annular crystalline membranes. The schematic plot of the model system is shown
in Fig.~\ref{sch}. The L-J lattice model consists of particles in triangular
lattice under the L-J potential:
\be
V(r)=4\epsilon_0 \big[ \lb \f{\sigma_0}{r}\rb^{12} - \lb \f{\sigma_0}{r}\rb^{6} \big],
\ee
where $r$ is the distance between two particles, the parameters $\sigma_0$ and
$\epsilon_0$ are related to the length scale and energy scale of the L-J
potential. The potential energy has the lowest value $-\epsilon_0$ at the
balance distance of $\ell_0=2^{1/6}\sigma_0$. In this work, $\ell_0$ is set to
be the unit of length. The L-J potential is featured with the energy minimum
structure in the potential curve, which enables the breaking of the bonds and
thus allows one to capture the physical process of the plastic
deformation~\cite{jones1924determination}. In simulations, a cut-off length of
$r_c=3\ell_0$ is introduced in the L-J potential.

The crystalline membrane consisting of L-J particles is confined in an annulus of inner
radius $\rho_{1}$ and outer radius $\rho_2$. In the initial configuration, the lattice
spacing is set to be the balance length $\ell_0$ of the L-J potential. The deformation of
the annular membrane is driven by gradually expanding the inner boundary. The outer
boundary consists of a few layers of regularly packed particles, as highlighted in
Fig.~\ref{sch}.  The lattice of these boundary particles is compatible with that of the
membrane to avoid the introduction of topological defects that may lead to uncontrollable
displacement of the particles near the outer boundary in the expansion process.  In
simulations, to expand the inner boundary, we introduce a circle as a geometric
constraint, and gradually increase the radius of this circle to push the particles
outward. The displacement of the particles is confined on the plane. The amount of the
expansion rate $h$ is set to be as small as $0.2\ell_0$ to fulfill the quasi-static
condition. The radius of the inner boundary is increased by $h$ in each expansion. The
new radius of the inner boundary is denoted as $\rho'_{1}$.  After each expansion, the
system is mechanically relaxed. By the standard steepest descend method, the local
lowest-energy particle configuration is identified at the resolution of the step size
$s=10^{-4}\ell_0$~\cite{snyman2005practical,yao2016electrostatics}.

\section{Results and discussion}

This section consists of two subsections. In Sec. III A, we first analyze the
small deformation of the annular membrane by the combination of analytical
elasticity theory and numerical simulations. The agreement of the numerical and
theoretical results shows the reliability of the computational approach, which
is used to explore the regime of plastic deformation. In Sec. III B, we explore
the phenomenon of plastic instability in the regime of large deformation. The
reorganization of the particles upon the gradual expansion of the inner boundary
is analyzed from the perspective of topological defect. We reveal the
characteristic dynamic structures of the dislocation current and the concurrent
vortices in the plastic deformation process.

\subsection{Elasticity analysis in the regime of small deformation}

\begin{figure}[t]  % h: put figure just in this position if possible
\centering 
\includegraphics[width=3in]{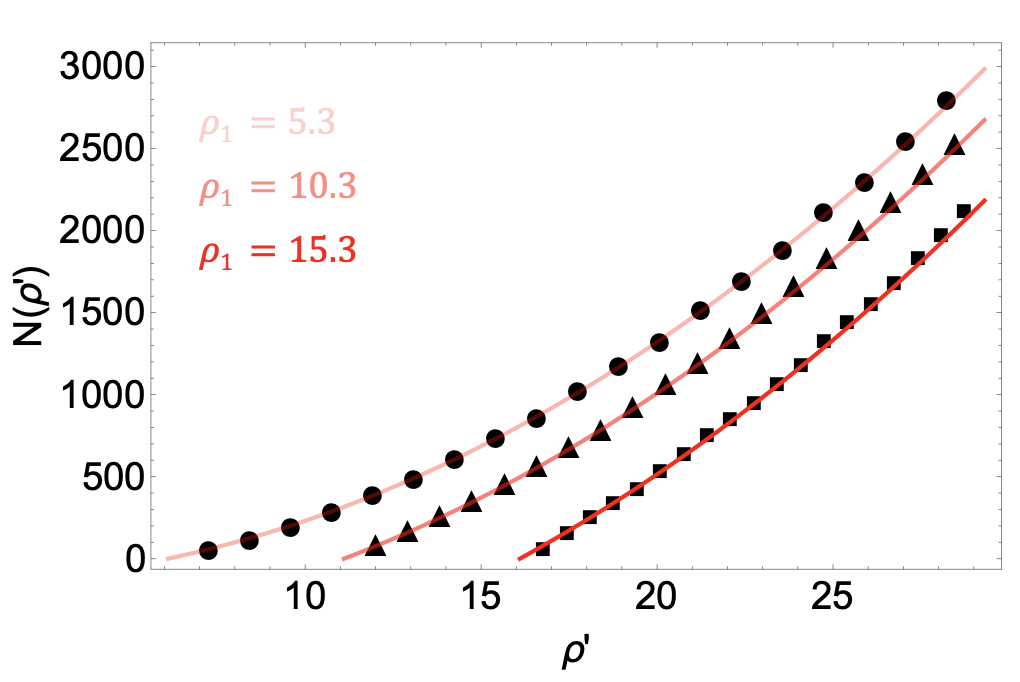}
  \caption{Cumulative particle distributions over the deformed annular membrane
  at varying initial radius of the inner boundary.  $N(\rho')$ is the total
  number of particles within the circle of radius $\rho'$ in the deformed
  lattice; $\rho'\in [\rho_1+h, \rho_2]$. The expansion rate $h =
  0.8$. The simulation data (dots) could be well fitted by the theoretical
  results (red curves). The radius of the outer boundary $\rho_2=29.4$.  }
\label{N_rho}
\end{figure}

% theory

We first resort to analytical continuum elasticity theory to analyze the 2D elastic
deformation of the annular membrane system under the radial expansion of the
inner circular boundary~\cite{Landau1986,geminard2004wrinkle}. For an isotropic membrane in
mechanical equilibrium, the displacement field $\vec{u}$ is governed by the
following balance equation in the absence of body force: 
\begin{eqnarray}
  \nabla \nabla \cdot \vec{u} - \frac{1}{2} (1-\sigma) \nabla \times  (\nabla
  \times \vec{u}) = 0, \label{balance}
\end{eqnarray}
where $\sigma$ is the Poisson's ratio. To discuss the case of radial expansion
of the inner boundary, we work in polar coordinates $(\rho, \varphi)$. The inner
circular boundary is expanded by an amount of $h$. The boundary conditions are:
$u_{\rho}(\rho=\rho_1)=h$, $u_{\varphi}(\rho=\rho_1)=0$,
$u_{\rho}(\rho=\rho_2)=0$, and $u_{\varphi}(\rho=\rho_2)=0$.

Consider the axisymmetric solution:
\begin{eqnarray}
  \vec{u}(\rho) = u_{\rho}(\rho) \hat{e}_{\rho}, \label{sol_B}
\end{eqnarray} 
where $\hat{e}_{\rho}$ is a unit vector along the radial direction.
By inserting Eq.~(\ref{sol_B}) into Eq.~(\ref{balance}), the balance equation
becomes
\begin{eqnarray}
  \partial_{\rho}[\frac{1}{\rho} \partial_{\rho} (\rho u_{\rho}) ] = 0.
  \label{u_rho_B1}
\end{eqnarray} 
Note that the effect of the Poisson ratio vanishes, because $\nabla \times
\vec{u}=0$ for the axisymmetric solution in Eq.~(\ref{sol_B}). By substituting the boundary conditions into
Eq.~(\ref{sol_B}), we obtain the following solution:
\begin{eqnarray}
  u_{\rho}  = \frac{h \rho_1}{\rho_2^2-\rho_1^2}\frac{\rho_2^2-\rho^2}{\rho}
  \label{u_rho_B2},
\end{eqnarray} 
where $\rho \in [\rho_1, \rho_2]$. One may check that $u_{\rho}(\rho_1) = h$ and
$u_{\rho}(\rho_2) = 0$ in Eq.~(\ref{u_rho_B2}). From Eq.~(\ref{u_rho_B2}), we
derive for the strain along the radial and azimuthal directions:
\begin{eqnarray}
  u_{\rho\rho}  &=& -\frac{h \rho_1}{\rho_2^2-\rho_1^2} (1+\frac{\rho_2^2}{\rho^2}), \nonumber \\
  u_{\varphi\varphi} &=& \frac{h \rho_1}{\rho_2^2-\rho_1^2}(\frac{\rho_2^2}{\rho^2}-1).
  \label{strain_B}
\end{eqnarray}

Now, we analyze the strain field in the annular membrane upon the expansion of
the inner boundary based on Eq.~(\ref{strain_B}). Overall, the magnitudes of
both $u_{\rho\rho}$ and $u_{\varphi\varphi}$ decrease in the form of $\sim
\rho^{-2}$ when approaching the outer boundary. The radial expansion of the
circular boundary originally at $\rho=\rho_1$ leads to radial compression
(negative $u_{\rho\rho}$) and azimuthal stretching (positive
$u_{\varphi\varphi}$). The signs of $u_{\rho\rho}$ and $u_{\varphi\varphi}$ are
always opposite over the entire annular membrane ($\rho \in [\rho_1,
\rho_2]$).  This observation has implications in the stability of the membrane.
In elastic membrane theory, it has been proved that an equilibrium configuration
cannot be stable or neutrally stable unless the principal stresses are
everywhere nonnegative~\cite{steigmann1986proof}. By this criterion, pulling
the inner boundary inward may lead to wrinkles in the region of $\rho/\rho_2 <
\rho^*$.  $\rho^* = \sqrt{(1-\sigma)/(1+\sigma)}$ ($\rho_1 >
\rho^*$)~\cite{geminard2004wrinkle,plaut2009linearly}.

%% compare simulation and theory

We proceed to perform numerical simulations to study the elastic deformation of the
L-J lattice model, and compare the numerical and analytical results. The goal is to
check the reliability of the computational approach, which will be used to
explore the regime of plastic deformation, and also to analyze the region of
validity of the linear elasticity theory for the L-J lattice model.

In order to compare with the theoretical results, we first count the total number of
particles $N(\rho')$ within the circle of radius $\rho'$ in the deformed annular
lattice in mechanical equilibrium; the prime symbol in
$\rho'$ is to indicate that the variable $\rho'$ is defined over the deformed
membrane. $\rho'$ ranges from $\rho_1+h$ to $\rho_2$. In the following, we
derive $N(\rho')$ from Eq.~(\ref{strain_B}).
First of all, the area element $dA$ in the undeformed lattice becomes $dA'$ in the
deformation. The relation of $dA'$ and $dA$ is:
\begin{eqnarray}
  \frac{dA'}{dA}  = (1+u_{\rho\rho})(1+u_{\varphi\varphi}).
  \label{dA_B}
\end{eqnarray} 
Note that in the regime of linear elasticity, Eq.(\ref{dA_B}) is written as the trace of the strain
tensor:
\begin{eqnarray}
  \frac{dA'}{dA}  \approx 1+u_{\rho\rho}+u_{\varphi\varphi}.
  \label{dA_B2}
\end{eqnarray} 
According to Eq.(\ref{dA_B}), the particle density defined on the undeformed lattice is changed
from $\lambda(\rho)$ to $\lambda'(\rho)$, where $\rho \in [\rho_1, \rho_2]$.
$\lambda$ to $\lambda'$ are related by
\begin{eqnarray}
  \frac{\lambda'}{\lambda}  = \frac{dA}{dA'} 
  =
  [(1+u_{\rho\rho})(1+u_{\varphi\varphi})]^{-1}.
  \label{density_B}
\end{eqnarray}
Equation (\ref{density_B}) clearly shows that stretching the lattice (positive
$u_{\rho\rho}$ and $u_{\varphi\varphi}$) leads to the reduction of the particle
density. The number of particles
within the circle of radius $\rho$ in the unformed lattice is as follows: 
\begin{eqnarray}
  N(\rho)  = \int_{\rho_1}^{\rho} \lambda(\rho) d\rho.
  \label{densityrho_B}
\end{eqnarray}

Now, we count the total number of particles $N(\rho')$ within the circle of
radius $\rho'$ from the deformed annular lattice in mechanical equilibrium. To
this end, we shall consider the movement of a particle from $\rho$ to $\rho'$ in
the deformation.  $\rho$ and $\rho'$ are connected by
\begin{eqnarray}
  \rho' = \rho + u_{\rho}.
  \label{rhop_B}
\end{eqnarray} 
The displacement vector $u_{\rho}$ is defined on the undeformed lattice. The
expression for $u_{\rho}$ is given in Eq.~(\ref{u_rho_B2}). 
The plots of $N(\rho')$ at varying initial radius of the inner circular boundary
($\rho_1$) are presented in Fig.~\ref{N_rho}.  The simulation data (dots of
different shapes) could be well fitted by the theoretical results (solid red
curves).  Simulations show that the results of simulation and linear elasticity
theory agree well up to $h/(\rho_2-\rho_1) \approx 10\%$ for the case of
$\rho_1=15.3$ and $\rho_2=29.4$.

\subsection{Plastic instability in the regime of large deformation}

\begin{figure*}[t]  % h: put figure just in this position if possible
\centering 
\includegraphics[width=7in]{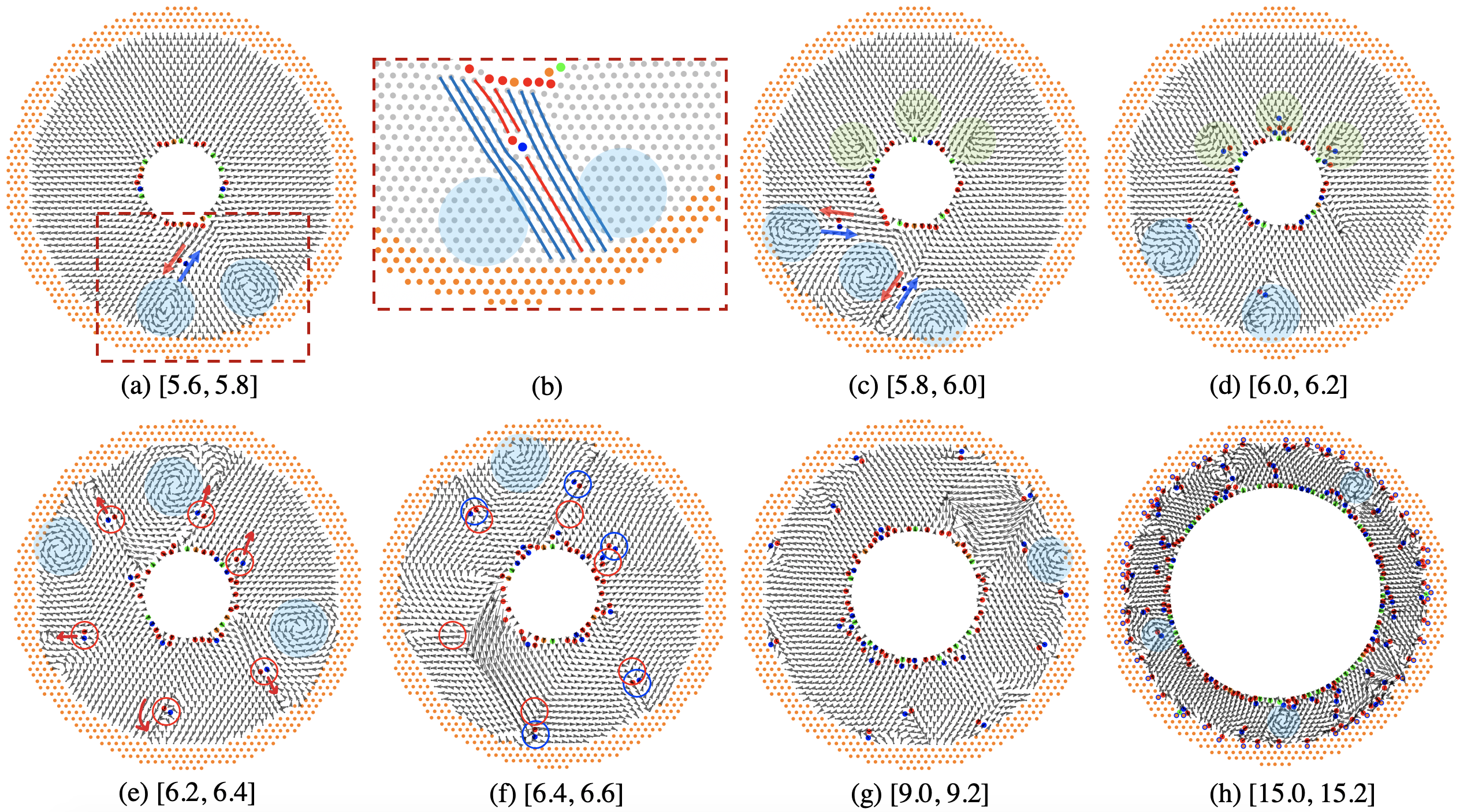}
  \caption{The pattern of dislocation current is revealed in the 2D plastic
  deformation of the annular crystalline membrane upon continuous expansion of
  the inner boundary. The displacement vectors of the particles as the radius
  of the inner boundary is increased from $\rho_i$ to $\rho_f$ are represented
  by the arrows; the values of $\rho_i$ and $\rho_f$ are given in the square
  bracket below each figure. (a) The red and blue arrows indicate the shearing
  deformation, leading to the emergence of the dislocation and the vortices
  (as indicated by the blue disks). The dislocation is composed of a pair of
  five- and seven-fold disclinations as represented by the red and blue dots.
  The boxed region in panel (a) is enlarged in panel (b) for showing the reorganization of
  the crystal lattice by the dislocation. With the expansion of the inner
  boundary, we observe the collective migration of the dislocations from the inner
  to the outer boundary, as shown in panels (e) and (f). In panel (f), the previous locations
  of the dislocations are indicated by red circles for visual convenience.  The
  system is stress-free in the initial state, where the value of the radius of
  the inner boundary $\rho_{1}=5$, and $\rho_2 = 24.4$. The annular L-J lattice
  consists of 1574 particles. }
\label{shear}
\end{figure*}

We resort to the computational approach to explore the regime of
large deformation, where the L-J lattice model exhibits plastic deformations. 
The reorganization of the particles in the strong expansion of the inner
boundary is analyzed from the perspectives of topological defect.

Figure~\ref{shear} shows the plastic deformations of the annular lattice with
the expansion of the inner circular boundary. The plastic deformation is
characterized by the emergence of disclinations, which are indicated by colored
dots in Fig.~\ref{shear}. Disclinations are identified by the standard
Delaunay triangulation~\cite{nelson2002defects}. The five- and seven-fold
disclinations are represented by the red and blue dots. A pair of five- and
seven-fold disclinations constitute a dislocation. To track the deformation
process, we also plot the displacement field associated with the expansion of
the inner boundary from $\rho'_1 = \rho_i$ to $\rho'_1 = \rho_f$; the values of
$\rho_i$ and $\rho_f$ are given in the square brackets below each figure.

As a signal for the occurrence of the plastic deformation, the first dislocation
appears when the radius of the inner boundary is increased from $\rho_1'=5$ to
$\rho_1'=5.6$, as shown in Fig.~\ref{shear}(a). The zoomed-in particle
configuration near the dislocation is presented Fig.~\ref{shear}(b). The
particle arrays indicated by the blue and red lines show the effect of the
dislocation on the reorganization of the particles upon the expansion.
Specifically, in the presence of the dislocation, an extra particle array along
the red line is introduced in the interior side of the annular lattice. The
appearance of this inserted particle array terminated at the inner boundary
reflects the adaptation of the particle configuration to the increased perimeter
of the inner boundary upon expansion.  The perspective of topological defect
allows one to capture this neat self-adaption process.

\begin{figure}[t]  % h: put figure just in this position if possible
\centering 
\includegraphics[width=2.3in]{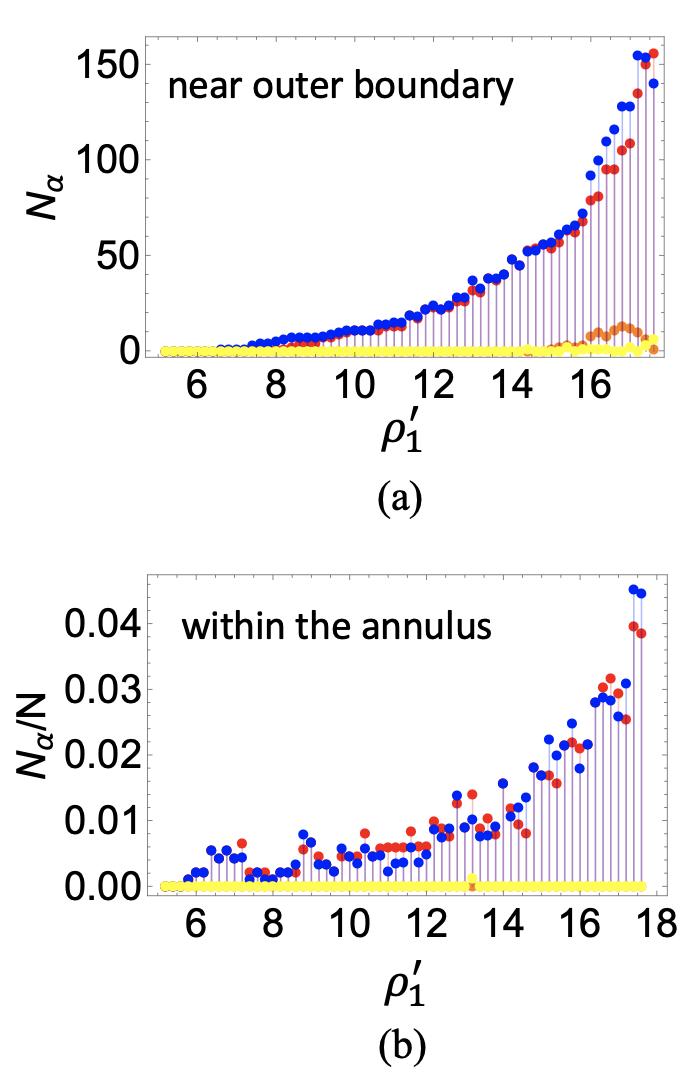}
  \caption{Statistics of the emergent disclinations in the plastic deformation
  of the annular membrane driven by the expansion of the inner boundary. (b)
  Plot of the number of $\alpha$-fold disclinations $N_{\alpha}$ accumulated
  near the outer boundary vs the inner radius $\rho'_1$ of the deformed
  membrane.  The defects within the thin annular region of $1.5$ lattice
  spacings away from the outer boundary are counted. (b) Plot of the relative
  number of $\alpha$-fold disclinations within the annular membrane vs the inner
  radius $\rho'_1$ of the deformed membrane.  The thin rims (1.5 lattice
  spacings) near both inner and outer boundaries are excluded. The four-, five-,
  seven-, and eight-fold disclinations are indicated by orange, red, blue, and
  yellow dots, respectively. In the initial stress-free state, the radius of the
  inner boundary $\rho_{1}=5$, and $\rho_2 = 24.4$. $N=1574$.   
    }
\label{defect_count}
\end{figure}

The question of why the dislocation appears upon the expansion of the inner
boundary is discussed in the preceding discussion. We further inquire how the
dislocation arises. To address this question, we carefully analyze the
displacement field associated with the incremental expansion of the inner
circular boundary, as shown in Fig.~\ref{shear}(a). Near the dislocation, we
find anti-parallel arrays of the displacement vectors (indicated by the pair of
red and blue arrows) that are perpendicular to the line connecting the 5- and
7-fold disclinations composing the dislocation. Such a pattern of the
displacement field is an indicator of the shearing deformation. As such, the
appearance of the dislocation is caused by the shearing deformation of particle
arrays. Note that the crystal instability phenomenon is conventionally
analyzed in terms of fluctuations of varying wavelength in the approach of
statistical field theory~\cite{born1940,Landau1999a}. Here, the computational
approach enables the analysis of the microscopic shearing process, which reveals the
detailed information on the crystal disruption process.

We further observe that the shearing deformation leads to the emergence of a
pair of clockwise and counterclockwise vortices as highlighted by the blue disks
in Fig.~\ref{shear}(a). In other words, the reorganization of the particles upon
the expansion involves the collective movement of the particles, where the
coherent vortex structure emerges in the displacement field. While the
shear-driven emergence of vortices is common in fluids, the observed vortex structure in
our solid mechanical system and its crucial role in reorganizing the particles
upon stress are remarkable. Fig.~\ref{shear}(b) shows that in the regions of the
vortices indicated by the blue disks, the crystalline order is well preserved.
Here, we shall point out that due to the matched lattices of the particle array
in the outer boundary and in the crystalline membrane, the entire annular region
near the outer boundary is free of defects in the expansion process until
$\rho_1'$ is increased from the initial value of $5$ to about $6.6$.

Further expanding the inner boundary leads to the emergence of the second
dislocation, as shown in Fig.~\ref{shear}(c). Simultaneously, an extra vortex
(the left one among the three indicated by blue disks) is observed. We further
notice the irregular distribution of the displacement vectors in the regions
indicated by green disks, where the crystalline order is still well preserved.
Upon further slight expansion as shown in Fig.~\ref{shear}(d), either
dislocations or compound defects appear in these irregular spots in the
displacement field. Therefore, the irregular spots in the displacement field
serve as precursors for the ensuing appearance of topological defects in the
lattice. Specifically, identifying these irregular spots allows us to pinpoint
the location of the occurring topological defects. Here, it is shown that
scrutiny of the displacement field yields useful insights into the plastic
deformation.

Here, it is of interest to discuss the dynamics of the vortices in the
displacement field with the expansion of the inner boundary. We observe the
generation of vortices accompanying the shearing deformation, as shown in
Figs.~\ref{shear}(a) and \ref{shear}(b). Comparison of Figs.~\ref{shear}(c) and
\ref{shear}(d) also shows the annihilation of a vortex. Furthermore, from
Figs.~\ref{shear}(a) to \ref{shear}(d), we see that a gentle expansion by a
fraction of one lattice spacing leads to a large displacement of the vortex
center at the order of ten lattice spacings, indicating the fast dynamics of the
vortices.

Under stronger expansion, we observe a proliferation of dislocations as shown in the
lower panels in Fig.~\ref{shear}. In the presence of these dislocations, the
entire displacement field is further distorted; more vortices and irregular
spots arise. The microscopic scenario of the plastic deformation of the
annular lattice exhibits complicated interplay of topological defects and
displacement field.

Is there any pattern underlying the plastic deformation process? To address this
question, we track the movement of the dislocations by comparing the particle
configurations in the gradual expansion of the inner boundary.  Simulations show
that the circled dislocations in Fig.~\ref{shear}(e) move to the locations in
the blue circles in Fig.~\ref{shear}(f), where the previous locations of the
dislocations are indicated by red circles for visual convenience. We also
observe the rotation of the dislocation, as shown in the lowest circled
dislocation in Fig.~\ref{shear}(e). The expansion-driven radial migration of
dislocations towards the outer boundary continues under stronger expansion.  As
such, the migrating dislocations form a radial current across the annular
membrane. The inner boundary that is filled with defects serves as the source,
and the outer boundary plays the role of the sink in the current of the
dislocations.  Upon their arrival at the outer boundary, the dislocations are
anchored therein, leading to the accumulation of dislocations at the outer
boundary in the expansion of the inner boundary. The plastic deformation process
in Fig.~\ref{shear} is recorded in a movie presented in the Supplemental
Material. Note that the dynamic scenario of the current of dislocations is
invariant under the finer expansion rates of $h=0.1\ell_0$ and $h=0.2\ell_0$ for the
case in Fig.~\ref{shear}.

The accumulation of topological defects at the outer boundary is quantitatively
analyzed. In Fig.~\ref{defect_count}(a), we show the variation of the number of
$\alpha$-fold disclinations ($N_{\alpha}$) near the outer boundary in the
expansion of the inner boundary. The radius of the inner boundary $\rho_{1}=5$
in the initial stress-free state. The defects within the region of 1.5 lattice
spacings away from the outer boundary are counted. The four-, five-, seven-, and
eight-fold disclinations are indicated by orange, red, blue, and yellow dots,
respectively. Figure~\ref{defect_count}(a) shows that the five- and seven-fold
disclinations dominate over the other types of disclinations.  We see that the
value of $N_{\alpha}$ becomes nonzero when $\rho_1$ exceeds about 6.6 for the
case of $\rho_{1}=5$. In other words, the first dislocation arrives at the outer
boundary when the amount of the expansion of the inner boundary exceeds some
critical value $\delta \rho_c$; $\delta \rho_c=1.6$ for the case of $\rho_{1}=5$
and $\rho_2=24.4$.  Simulations show that the value of $\delta \rho_c$ is
insensitive to the initial radii of the inner and outer boundaries.
Specifically, the value of $\delta \rho_c$ is in the range of $1.4\pm 0.2$ for
$\rho_{1} = \{5, 10, 15 \}$ and $\rho_2=\{20.4, 24.4, 28.4, 32.4 \}$.

Here, we shall point out that the geometric incompatibility of the triangular
lattice in the crystalline membrane and the circular inner boundary leads to the
formation of disclinations within the annular region near the inner boundary. While the
dislocations migrating to the outer boundary may be regarded as being pulled out of the
crowd of these disclinations, the absence of isolated dislocations therein makes
it difficult to describe their nucleation analytically. Especially, it is a
challenge to analytically determine the critical value of $\rho_1'$, above which
the first batch of dislocations start to move out of the crowded disclinations
at the inner boundary.  However, the common value of $\delta \rho_c$ in systems
of varying geometry in the preceding discussion gives an upper bound estimate
for the critical value of $\rho_1'$.

In Fig.~\ref{defect_count}(b), we also show the expansion-driven variation of
the relative number of $\alpha$-fold disclinations $N_{\alpha}$ within the
annulus of the L-J lattice. Note that in the counting of the disclinations, the
thin rims (1.5 lattice spacings) near both inner and outer boundaries are
excluded.  Similar to the case in Fig.~\ref{defect_count}(a), the five- and
seven-fold disclinations (indicated by red and blue dots) dominate over the
other types of disclinations. Figure~\ref{defect_count}(b) shows that the
proportion of defects is only about $4\%$ even when the value of $\rho_1$ is
enlarged by over three times. The small amount of topological defects in the
interior of the annular membrane suggests that the current of dislocations
provides the mechanism for protecting the crystalline order. This result may
have implication to the restoration of the crystalline order by creating a flow
of disclinations in 2D packings of particles.

In the preceding discussions, we reveal the scenario of the current of dislocations
underlying the expansion-driven plastic deformation of the annular lattice. It
is of interest to inquire about the generality of this physical scenario. We perform
further simulations for the elliptic system, where the lattice is confined
between an inner circular boundary and an outer elliptic boundary. The current
of dislocations is also observed in the elliptic system. The disruption of the
crystalline order in the elliptic system upon the expansion of the inner
boundary is recorded in a movie presented in the Supplemental Material.

Migration of dislocations has also been reported in the transport of interacting
vortices in the superconducting film of the Corbino disk
geometry~\cite{natmatcarmen2003,miguel2011laminar}. The motion of the vortices
is driven by an electric current that is injected at the center of the disk
and flows radially towards the boundary. With the increase of the current, the
vortex array that is initially in triangular lattice experiences a tearing
transition from rigid rotation to plastic flow as characterized by the
nucleation of dislocations at the center of the disk.  Under a stronger
current, the dislocations are organized into radial grain boundaries, coherently
gliding in the tangential direction and forming a laminar flow over the
disk~\cite{miguel2011laminar}.

Here, we compare these two kinds of systems: the vortex array in Corbino disk
geometry and the crystalline membrane in annular
geometry~\cite{natmatcarmen2003,miguel2011laminar}. The vortex array and the
electric current in the former system are analogous to the L-J lattice and the
mechanical expansion in the latter system, respectively. Both studies focus on
the plastic behavior of regularly packed elementary constituents in response to
an external driving force of increasing strength.  Despite of the distinct
physical settings, both systems share a common topological scenario in terms of
the migration of dislocations in the plasticity phenomena. The topological
perspective based on the dynamics of dislocations offers a unified frame to
understand nonequilibrium collective properties of a host of ordered
self-assemble structures.  Regarding the difference of the two kinds of systems,
the Corbino disk system exhibits a second transition to the formation of the
coherent laminar flow along the tangential direction under an even stronger
current~\cite{miguel2011laminar}; such a transition is absent in the mechanical
membrane system due to the lack of a tangential force on the particles.
Furthermore, in the Corbino disk system, the Lorentz force caused by the imposed
external current is applied to each vortex, and the direction of the force is
azimuthal.  In contrast, the external force is applied to a single layer of
particles at the inner boundary in the mechanical membrane system, and the force
is along the radial direction.

We finally briefly discuss the possible extensions of the current work. First,
one may extend the planar geometry, where the 2D crystalline membrane lives, to
curved surfaces. The geometric frustration of the crystalline order on various
curved surfaces has been well studied in the regime of mechanical equilibrium,
and a series of defect motifs have been
revealed~\cite{bowick2009two,wales2014chemistry}. It is of interest to utilize
the stress focusing effect around the pre-existent topological defects in curved
crystals to regulate the patterns of plastic instability upon external
mechanical
stimuli~\cite{cottrell1965dislocations,nelson2002defects,chen2022geometry}.
Second, the quasi-static process, on which the current work focuses, may be
extended to the dynamical regime by applying temporally-varying external
stress~\cite{negri2015deformation}. The dynamical annular membrane system
provides a suitable model to explore the interplay of topological defects and
elastic waves, which has potential connections to the control of the dislocation
current as revealed in this work and that of the plastic instability process.

\section{Conclusion}

In summary, we study the plastic instability of the crystalline membrane system
in annular geometry. The expansion of the inner boundary leads to the
reorganization of the particles, which is analyzed from the perspective of
topological defect. We reveal the dynamic pattern of the plastic deformation in
the form of the dislocation current. The continuously generated dislocations at
the inner boundary collectively migrate to the outer boundary.  We also identify
the characteristic vortex structure arising from the interplay of topological
defects and the displacement field in the process of plastic deformation. These
results may find applications in the precise control of structural instabilities
in packings of particles (colloids, proteins, etc.,) and covalently
bonded systems.

\section{Acknowledgements}

This work was supported by the National Natural Science Foundation of China
(Grants No. BC4190050).

%\bibliography{/Users/zyao/Documents/Researches/jabref/ref_jab_ONE}

\begin{thebibliography}{53}
\expandafter\ifx\csname natexlab\endcsname\relax\def\natexlab#1{#1}\fi
\expandafter\ifx\csname bibnamefont\endcsname\relax
  \def\bibnamefont#1{#1}\fi
\expandafter\ifx\csname bibfnamefont\endcsname\relax
  \def\bibfnamefont#1{#1}\fi
\expandafter\ifx\csname citenamefont\endcsname\relax
  \def\citenamefont#1{#1}\fi
\expandafter\ifx\csname url\endcsname\relax
  \def\url#1{\texttt{#1}}\fi
\expandafter\ifx\csname urlprefix\endcsname\relax\def\urlprefix{URL }\fi
\providecommand{\bibinfo}[2]{#2}
\providecommand{\eprint}[2][]{\url{#2}}

\bibitem[{\citenamefont{Ghoniem and Walgraef}(2008)}]{Instabilities2008}
\bibinfo{author}{\bibfnamefont{N.}~\bibnamefont{Ghoniem}} \bibnamefont{and}
  \bibinfo{author}{\bibfnamefont{D.}~\bibnamefont{Walgraef}},
  \emph{\bibinfo{title}{{Instabilities and Self-Organization in Materials}}}
  (\bibinfo{publisher}{Oxford University Press}, \bibinfo{year}{2008}).

\bibitem[{\citenamefont{Audoly and Pomeau}(2010)}]{audoly2010elasticity}
\bibinfo{author}{\bibfnamefont{B.}~\bibnamefont{Audoly}} \bibnamefont{and}
  \bibinfo{author}{\bibfnamefont{Y.}~\bibnamefont{Pomeau}},
  \emph{\bibinfo{title}{Elasticity and Geometry}} (\bibinfo{publisher}{Oxford
  University Press}, \bibinfo{year}{2010}).

\bibitem[{\citenamefont{Bostwick et~al.}(2016)\citenamefont{Bostwick, Miksis,
  and Davis}}]{bostwick2016elastic}
\bibinfo{author}{\bibfnamefont{J.}~\bibnamefont{Bostwick}},
  \bibinfo{author}{\bibfnamefont{M.}~\bibnamefont{Miksis}}, \bibnamefont{and}
  \bibinfo{author}{\bibfnamefont{S.}~\bibnamefont{Davis}}, \bibinfo{journal}{J.
  Roy. Soc. Interface} \textbf{\bibinfo{volume}{13}}, \bibinfo{pages}{20160408}
  (\bibinfo{year}{2016}).

\bibitem[{\citenamefont{Cerda and Mahadevan}(2003)}]{cerda2003geometry}
\bibinfo{author}{\bibfnamefont{E.}~\bibnamefont{Cerda}} \bibnamefont{and}
  \bibinfo{author}{\bibfnamefont{L.}~\bibnamefont{Mahadevan}},
  \bibinfo{journal}{Phys. Rev. Lett.} \textbf{\bibinfo{volume}{90}},
  \bibinfo{pages}{074302} (\bibinfo{year}{2003}).

\bibitem[{\citenamefont{Marder et~al.}(2007)\citenamefont{Marder, Deegan, and
  Sharon}}]{Marder2007}
\bibinfo{author}{\bibfnamefont{M.}~\bibnamefont{Marder}},
  \bibinfo{author}{\bibfnamefont{R.}~\bibnamefont{Deegan}}, \bibnamefont{and}
  \bibinfo{author}{\bibfnamefont{E.}~\bibnamefont{Sharon}},
  \bibinfo{journal}{Physics Today}  (\bibinfo{year}{2007}).

\bibitem[{\citenamefont{Li et~al.}(2012)\citenamefont{Li, Cao, Feng, and
  Gao}}]{li2012mechanics}
\bibinfo{author}{\bibfnamefont{B.}~\bibnamefont{Li}},
  \bibinfo{author}{\bibfnamefont{Y.-P.} \bibnamefont{Cao}},
  \bibinfo{author}{\bibfnamefont{X.-Q.} \bibnamefont{Feng}}, \bibnamefont{and}
  \bibinfo{author}{\bibfnamefont{H.}~\bibnamefont{Gao}}, \bibinfo{journal}{Soft
  Matter} \textbf{\bibinfo{volume}{8}}, \bibinfo{pages}{5728}
  (\bibinfo{year}{2012}).

\bibitem[{\citenamefont{Chopin and Kudrolli}(2013)}]{PhysRevLett.111.174302}
\bibinfo{author}{\bibfnamefont{J.}~\bibnamefont{Chopin}} \bibnamefont{and}
  \bibinfo{author}{\bibfnamefont{A.}~\bibnamefont{Kudrolli}},
  \bibinfo{journal}{Phys. Rev. Lett.} \textbf{\bibinfo{volume}{111}},
  \bibinfo{pages}{174302} (\bibinfo{year}{2013}).

\bibitem[{\citenamefont{Kantor et~al.}(1986)\citenamefont{Kantor, Kardar, and
  Nelson}}]{kantor1986statistical}
\bibinfo{author}{\bibfnamefont{Y.}~\bibnamefont{Kantor}},
  \bibinfo{author}{\bibfnamefont{M.}~\bibnamefont{Kardar}}, \bibnamefont{and}
  \bibinfo{author}{\bibfnamefont{D.~R.} \bibnamefont{Nelson}},
  \bibinfo{journal}{Phys. Rev. Le} \textbf{\bibinfo{volume}{57}},
  \bibinfo{pages}{791} (\bibinfo{year}{1986}).

\bibitem[{\citenamefont{Lobkovsky et~al.}(1995)\citenamefont{Lobkovsky,
  Gentges, Li, Morse, and Witten}}]{lobkovsky1995scaling}
\bibinfo{author}{\bibfnamefont{A.}~\bibnamefont{Lobkovsky}},
  \bibinfo{author}{\bibfnamefont{S.}~\bibnamefont{Gentges}},
  \bibinfo{author}{\bibfnamefont{H.}~\bibnamefont{Li}},
  \bibinfo{author}{\bibfnamefont{D.}~\bibnamefont{Morse}}, \bibnamefont{and}
  \bibinfo{author}{\bibfnamefont{T.~A.} \bibnamefont{Witten}},
  \bibinfo{journal}{Science} \textbf{\bibinfo{volume}{270}},
  \bibinfo{pages}{1482} (\bibinfo{year}{1995}).

\bibitem[{\citenamefont{Nelson et~al.}(2004)\citenamefont{Nelson, Piran, and
  Weinberg}}]{Nelson2004c}
\bibinfo{author}{\bibfnamefont{D.~R.} \bibnamefont{Nelson}},
  \bibinfo{author}{\bibfnamefont{T.}~\bibnamefont{Piran}}, \bibnamefont{and}
  \bibinfo{author}{\bibfnamefont{S.}~\bibnamefont{Weinberg}},
  \emph{\bibinfo{title}{Statistical Mechanics of Membranes and Surfaces}}
  (\bibinfo{publisher}{World Scientific, Singapore}, \bibinfo{year}{2004}).

\bibitem[{\citenamefont{Chen and Santangelo}(2010)}]{chen2010minimal}
\bibinfo{author}{\bibfnamefont{B.~G.-g.} \bibnamefont{Chen}} \bibnamefont{and}
  \bibinfo{author}{\bibfnamefont{C.~D.} \bibnamefont{Santangelo}},
  \bibinfo{journal}{Phys. Rev. E} \textbf{\bibinfo{volume}{82}},
  \bibinfo{pages}{056601} (\bibinfo{year}{2010}).

\bibitem[{\citenamefont{Balankin et~al.}(2010)\citenamefont{Balankin, Ochoa,
  Miguel, Ortiz, and Cruz}}]{balankin2010fractal}
\bibinfo{author}{\bibfnamefont{A.~S.} \bibnamefont{Balankin}},
  \bibinfo{author}{\bibfnamefont{D.~S.} \bibnamefont{Ochoa}},
  \bibinfo{author}{\bibfnamefont{I.~A.} \bibnamefont{Miguel}},
  \bibinfo{author}{\bibfnamefont{J.~P.} \bibnamefont{Ortiz}}, \bibnamefont{and}
  \bibinfo{author}{\bibfnamefont{M.~{\'A}.~M.} \bibnamefont{Cruz}},
  \bibinfo{journal}{Phys. Rev. E} \textbf{\bibinfo{volume}{81}},
  \bibinfo{pages}{061126} (\bibinfo{year}{2010}).

\bibitem[{\citenamefont{Davidovitch et~al.}(2011)\citenamefont{Davidovitch,
  Schroll, Vella, Adda-Bedia, and Cerda}}]{Davidovitch2011}
\bibinfo{author}{\bibfnamefont{B.}~\bibnamefont{Davidovitch}},
  \bibinfo{author}{\bibfnamefont{R.~D.} \bibnamefont{Schroll}},
  \bibinfo{author}{\bibfnamefont{D.}~\bibnamefont{Vella}},
  \bibinfo{author}{\bibfnamefont{M.}~\bibnamefont{Adda-Bedia}},
  \bibnamefont{and} \bibinfo{author}{\bibfnamefont{E.~A.} \bibnamefont{Cerda}},
  \bibinfo{journal}{Proc. Natl. Acad. Sci. U.S.A.}
  \textbf{\bibinfo{volume}{108}}, \bibinfo{pages}{18227}
  (\bibinfo{year}{2011}).

\bibitem[{\citenamefont{King et~al.}(2012)\citenamefont{King, Schroll,
  Davidovitch, and Menon}}]{king2012elastic}
\bibinfo{author}{\bibfnamefont{H.}~\bibnamefont{King}},
  \bibinfo{author}{\bibfnamefont{R.~D.} \bibnamefont{Schroll}},
  \bibinfo{author}{\bibfnamefont{B.}~\bibnamefont{Davidovitch}},
  \bibnamefont{and} \bibinfo{author}{\bibfnamefont{N.}~\bibnamefont{Menon}},
  \bibinfo{journal}{Proc. Natl. Acad. Sci. U.S.A.}
  \textbf{\bibinfo{volume}{109}}, \bibinfo{pages}{9716} (\bibinfo{year}{2012}).

\bibitem[{\citenamefont{Grason and Davidovitch}(2013)}]{Grason2013}
\bibinfo{author}{\bibfnamefont{G.~M.} \bibnamefont{Grason}} \bibnamefont{and}
  \bibinfo{author}{\bibfnamefont{B.}~\bibnamefont{Davidovitch}},
  \bibinfo{journal}{Proc. Natl. Acad. Sci. U.S.A.}
  \textbf{\bibinfo{volume}{110}}, \bibinfo{pages}{12893}
  (\bibinfo{year}{2013}).

\bibitem[{\citenamefont{Gomes et~al.}(1989)\citenamefont{Gomes, Jyh, Ren,
  Rodrigues, and Furtado}}]{gomes1989mechanically}
\bibinfo{author}{\bibfnamefont{M.}~\bibnamefont{Gomes}},
  \bibinfo{author}{\bibfnamefont{T.}~\bibnamefont{Jyh}},
  \bibinfo{author}{\bibfnamefont{T.}~\bibnamefont{Ren}},
  \bibinfo{author}{\bibfnamefont{I.}~\bibnamefont{Rodrigues}},
  \bibnamefont{and} \bibinfo{author}{\bibfnamefont{C.}~\bibnamefont{Furtado}},
  \bibinfo{journal}{J. Phys. D: Appl. Phys.} \textbf{\bibinfo{volume}{22}},
  \bibinfo{pages}{1217} (\bibinfo{year}{1989}).

\bibitem[{\citenamefont{Fokker et~al.}(2019)\citenamefont{Fokker, Janbaz, and
  Zadpoor}}]{fokker2019crumpling}
\bibinfo{author}{\bibfnamefont{M.}~\bibnamefont{Fokker}},
  \bibinfo{author}{\bibfnamefont{S.}~\bibnamefont{Janbaz}}, \bibnamefont{and}
  \bibinfo{author}{\bibfnamefont{A.}~\bibnamefont{Zadpoor}},
  \bibinfo{journal}{RSC Adv.} \textbf{\bibinfo{volume}{9}},
  \bibinfo{pages}{5174} (\bibinfo{year}{2019}).

\bibitem[{\citenamefont{Ben~Amar and Pomeau}(1997)}]{ben1997crumpled}
\bibinfo{author}{\bibfnamefont{M.}~\bibnamefont{Ben~Amar}} \bibnamefont{and}
  \bibinfo{author}{\bibfnamefont{Y.}~\bibnamefont{Pomeau}},
  \bibinfo{journal}{Proc. Math. Phys. Eng. Sci.}
  \textbf{\bibinfo{volume}{453}}, \bibinfo{pages}{729} (\bibinfo{year}{1997}).

\bibitem[{\citenamefont{Cerda and Mahadevan}(1998)}]{cerda1998conical}
\bibinfo{author}{\bibfnamefont{E.}~\bibnamefont{Cerda}} \bibnamefont{and}
  \bibinfo{author}{\bibfnamefont{L.}~\bibnamefont{Mahadevan}},
  \bibinfo{journal}{Phys. Rev. Lett.} \textbf{\bibinfo{volume}{80}},
  \bibinfo{pages}{2358} (\bibinfo{year}{1998}).

\bibitem[{\citenamefont{Mora and Boudaoud}(2002)}]{mora2002thin}
\bibinfo{author}{\bibfnamefont{T.}~\bibnamefont{Mora}} \bibnamefont{and}
  \bibinfo{author}{\bibfnamefont{A.}~\bibnamefont{Boudaoud}},
  \bibinfo{journal}{Europhys. Lett.} \textbf{\bibinfo{volume}{59}},
  \bibinfo{pages}{41} (\bibinfo{year}{2002}).

\bibitem[{\citenamefont{Cerda and Mahadevan}(2005)}]{cerda2005confined}
\bibinfo{author}{\bibfnamefont{E.}~\bibnamefont{Cerda}} \bibnamefont{and}
  \bibinfo{author}{\bibfnamefont{L.}~\bibnamefont{Mahadevan}},
  \bibinfo{journal}{Proc. Math. Phys. Eng. Sci.}
  \textbf{\bibinfo{volume}{461}}, \bibinfo{pages}{671} (\bibinfo{year}{2005}).

\bibitem[{\citenamefont{Lobkovsky and Witten}(1997)}]{lobkovsky1997properties}
\bibinfo{author}{\bibfnamefont{A.~E.} \bibnamefont{Lobkovsky}}
  \bibnamefont{and} \bibinfo{author}{\bibfnamefont{T.}~\bibnamefont{Witten}},
  \bibinfo{journal}{Phys. Rev. E} \textbf{\bibinfo{volume}{55}},
  \bibinfo{pages}{1577} (\bibinfo{year}{1997}).

\bibitem[{\citenamefont{Matan et~al.}(2002)\citenamefont{Matan, Williams,
  Witten, and Nagel}}]{matan2002crumpling}
\bibinfo{author}{\bibfnamefont{K.}~\bibnamefont{Matan}},
  \bibinfo{author}{\bibfnamefont{R.~B.} \bibnamefont{Williams}},
  \bibinfo{author}{\bibfnamefont{T.~A.} \bibnamefont{Witten}},
  \bibnamefont{and} \bibinfo{author}{\bibfnamefont{S.~R.} \bibnamefont{Nagel}},
  \bibinfo{journal}{Phys. Rev. Lett.} \textbf{\bibinfo{volume}{88}},
  \bibinfo{pages}{076101} (\bibinfo{year}{2002}).

\bibitem[{\citenamefont{Vliegenthart and
  Gompper}(2006)}]{vliegenthart2006forced}
\bibinfo{author}{\bibfnamefont{G.}~\bibnamefont{Vliegenthart}}
  \bibnamefont{and} \bibinfo{author}{\bibfnamefont{G.}~\bibnamefont{Gompper}},
  \bibinfo{journal}{Nat. Mater.} \textbf{\bibinfo{volume}{5}},
  \bibinfo{pages}{216} (\bibinfo{year}{2006}).

\bibitem[{\citenamefont{Tallinen et~al.}(2008)\citenamefont{Tallinen,
  {\AA}str{\"o}m, and Timonen}}]{tallinen2008deterministic}
\bibinfo{author}{\bibfnamefont{T.}~\bibnamefont{Tallinen}},
  \bibinfo{author}{\bibfnamefont{J.~A.} \bibnamefont{{\AA}str{\"o}m}},
  \bibnamefont{and} \bibinfo{author}{\bibfnamefont{J.}~\bibnamefont{Timonen}},
  \bibinfo{journal}{Phys. Rev. Lett.} \textbf{\bibinfo{volume}{101}},
  \bibinfo{pages}{106101} (\bibinfo{year}{2008}).

\bibitem[{\citenamefont{Gim{\'e}nez-Ribes
  et~al.}(2023)\citenamefont{Gim{\'e}nez-Ribes, Motaghian, van~der Linden, and
  Habibi}}]{gimenez2023crumpled}
\bibinfo{author}{\bibfnamefont{G.}~\bibnamefont{Gim{\'e}nez-Ribes}},
  \bibinfo{author}{\bibfnamefont{M.}~\bibnamefont{Motaghian}},
  \bibinfo{author}{\bibfnamefont{E.}~\bibnamefont{van~der Linden}},
  \bibnamefont{and} \bibinfo{author}{\bibfnamefont{M.}~\bibnamefont{Habibi}},
  \bibinfo{journal}{Materials \& Design} \textbf{\bibinfo{volume}{232}},
  \bibinfo{pages}{112159} (\bibinfo{year}{2023}).

\bibitem[{\citenamefont{Cottrell}(1965)}]{cottrell1965dislocations}
\bibinfo{author}{\bibfnamefont{A.~H.} \bibnamefont{Cottrell}},
  \emph{\bibinfo{title}{Dislocations and Plastic Flow in Crystals}}
  (\bibinfo{publisher}{Clarendon Press}, \bibinfo{year}{1965}).

\bibitem[{\citenamefont{Fleck et~al.}(1994)\citenamefont{Fleck, Muller, Ashby,
  and Hutchinson}}]{fleck1994strain}
\bibinfo{author}{\bibfnamefont{N.}~\bibnamefont{Fleck}},
  \bibinfo{author}{\bibfnamefont{G.}~\bibnamefont{Muller}},
  \bibinfo{author}{\bibfnamefont{M.~F.} \bibnamefont{Ashby}}, \bibnamefont{and}
  \bibinfo{author}{\bibfnamefont{J.~W.} \bibnamefont{Hutchinson}},
  \bibinfo{journal}{Acta. Mater.} \textbf{\bibinfo{volume}{42}},
  \bibinfo{pages}{475} (\bibinfo{year}{1994}).

\bibitem[{\citenamefont{Nicola et~al.}(2006)\citenamefont{Nicola, Xiang,
  Vlassak, Van~der Giessen, and Needleman}}]{nicola2006plastic}
\bibinfo{author}{\bibfnamefont{L.}~\bibnamefont{Nicola}},
  \bibinfo{author}{\bibfnamefont{Y.}~\bibnamefont{Xiang}},
  \bibinfo{author}{\bibfnamefont{J.~J.} \bibnamefont{Vlassak}},
  \bibinfo{author}{\bibfnamefont{E.}~\bibnamefont{Van~der Giessen}},
  \bibnamefont{and}
  \bibinfo{author}{\bibfnamefont{A.}~\bibnamefont{Needleman}},
  \bibinfo{journal}{J. Mech. Phys. Solids} \textbf{\bibinfo{volume}{54}},
  \bibinfo{pages}{2089} (\bibinfo{year}{2006}).

\bibitem[{\citenamefont{Negri et~al.}(2015)\citenamefont{Negri, Sellerio,
  Zapperi, and Miguel}}]{negri2015deformation}
\bibinfo{author}{\bibfnamefont{C.}~\bibnamefont{Negri}},
  \bibinfo{author}{\bibfnamefont{A.~L.} \bibnamefont{Sellerio}},
  \bibinfo{author}{\bibfnamefont{S.}~\bibnamefont{Zapperi}}, \bibnamefont{and}
  \bibinfo{author}{\bibfnamefont{M.-C.} \bibnamefont{Miguel}},
  \bibinfo{journal}{Proc. Natl. Acad. Sci.} \textbf{\bibinfo{volume}{112}},
  \bibinfo{pages}{14545} (\bibinfo{year}{2015}).

\bibitem[{\citenamefont{Chen and Yao}(2022)}]{chen2022geometry}
\bibinfo{author}{\bibfnamefont{J.}~\bibnamefont{Chen}} \bibnamefont{and}
  \bibinfo{author}{\bibfnamefont{Z.}~\bibnamefont{Yao}}, \bibinfo{journal}{Soft
  Matter} \textbf{\bibinfo{volume}{18}}, \bibinfo{pages}{5323}
  (\bibinfo{year}{2022}).

\bibitem[{\citenamefont{Weil and Newmark}(1955)}]{weil1955large}
\bibinfo{author}{\bibfnamefont{N.}~\bibnamefont{Weil}} \bibnamefont{and}
  \bibinfo{author}{\bibfnamefont{N.}~\bibnamefont{Newmark}},
  \bibinfo{journal}{J. Appl. Mech.} \textbf{\bibinfo{volume}{22(4):}},
  \bibinfo{pages}{533} (\bibinfo{year}{1955}).

\bibitem[{\citenamefont{Chopin and Kudrolli}(2016)}]{chopin2016disclinations}
\bibinfo{author}{\bibfnamefont{J.}~\bibnamefont{Chopin}} \bibnamefont{and}
  \bibinfo{author}{\bibfnamefont{A.}~\bibnamefont{Kudrolli}},
  \bibinfo{journal}{Soft Matter} \textbf{\bibinfo{volume}{12}},
  \bibinfo{pages}{4457} (\bibinfo{year}{2016}).

\bibitem[{\citenamefont{Celli et~al.}(2020)\citenamefont{Celli, Lamaro,
  McMahan, Bordeenithikasem, Hofmann, and Daraio}}]{celli2020compliant}
\bibinfo{author}{\bibfnamefont{P.}~\bibnamefont{Celli}},
  \bibinfo{author}{\bibfnamefont{A.}~\bibnamefont{Lamaro}},
  \bibinfo{author}{\bibfnamefont{C.}~\bibnamefont{McMahan}},
  \bibinfo{author}{\bibfnamefont{P.}~\bibnamefont{Bordeenithikasem}},
  \bibinfo{author}{\bibfnamefont{D.}~\bibnamefont{Hofmann}}, \bibnamefont{and}
  \bibinfo{author}{\bibfnamefont{C.}~\bibnamefont{Daraio}},
  \bibinfo{journal}{J. Mech. Phys. Solids} \textbf{\bibinfo{volume}{145}},
  \bibinfo{pages}{104129} (\bibinfo{year}{2020}).

\bibitem[{\citenamefont{Nelson and Peliti}(1987)}]{Nelson1987}
\bibinfo{author}{\bibfnamefont{D.}~\bibnamefont{Nelson}} \bibnamefont{and}
  \bibinfo{author}{\bibfnamefont{L.}~\bibnamefont{Peliti}},
  \bibinfo{journal}{J. Phys. (France)} \textbf{\bibinfo{volume}{48}},
  \bibinfo{pages}{1085} (\bibinfo{year}{1987}).

\bibitem[{\citenamefont{Nelson}(2002)}]{nelson2002defects}
\bibinfo{author}{\bibfnamefont{D.~R.} \bibnamefont{Nelson}},
  \emph{\bibinfo{title}{Defects and Geometry in Condensed Matter Physics}}
  (\bibinfo{publisher}{Cambridge University Press, Cambridge},
  \bibinfo{year}{2002}).

\bibitem[{\citenamefont{Bowick and Giomi}(2009)}]{bowick2009two}
\bibinfo{author}{\bibfnamefont{M.~J.} \bibnamefont{Bowick}} \bibnamefont{and}
  \bibinfo{author}{\bibfnamefont{L.}~\bibnamefont{Giomi}},
  \bibinfo{journal}{Adv. Phys.} \textbf{\bibinfo{volume}{58}},
  \bibinfo{pages}{449} (\bibinfo{year}{2009}).

\bibitem[{\citenamefont{Wales}(2014)}]{wales2014chemistry}
\bibinfo{author}{\bibfnamefont{D.~J.} \bibnamefont{Wales}},
  \bibinfo{journal}{ACS Nano} \textbf{\bibinfo{volume}{8}},
  \bibinfo{pages}{1081} (\bibinfo{year}{2014}).

\bibitem[{\citenamefont{Jones}(1924)}]{jones1924determination}
\bibinfo{author}{\bibfnamefont{J.~E.} \bibnamefont{Jones}},
  \bibinfo{journal}{Proc. R. Soc. London, Ser. A.}
  \textbf{\bibinfo{volume}{106}}, \bibinfo{pages}{463} (\bibinfo{year}{1924}).

\bibitem[{\citenamefont{G{\'e}minard et~al.}(2004)\citenamefont{G{\'e}minard,
  Bernal, and Melo}}]{geminard2004wrinkle}
\bibinfo{author}{\bibfnamefont{J.-C.} \bibnamefont{G{\'e}minard}},
  \bibinfo{author}{\bibfnamefont{R.}~\bibnamefont{Bernal}}, \bibnamefont{and}
  \bibinfo{author}{\bibfnamefont{F.}~\bibnamefont{Melo}},
  \bibinfo{journal}{Eur. Phys. J. E} \textbf{\bibinfo{volume}{15}},
  \bibinfo{pages}{117} (\bibinfo{year}{2004}).

\bibitem[{\citenamefont{Plaut}(2009)}]{plaut2009linearly}
\bibinfo{author}{\bibfnamefont{R.}~\bibnamefont{Plaut}},
  \bibinfo{journal}{Acta. Mech.} \textbf{\bibinfo{volume}{202}},
  \bibinfo{pages}{79} (\bibinfo{year}{2009}).

\bibitem[{\citenamefont{Zhang et~al.}(2012)\citenamefont{Zhang, Duan, and
  Wang}}]{zhang2012tunable}
\bibinfo{author}{\bibfnamefont{Z.}~\bibnamefont{Zhang}},
  \bibinfo{author}{\bibfnamefont{W.}~\bibnamefont{Duan}}, \bibnamefont{and}
  \bibinfo{author}{\bibfnamefont{C.~M.} \bibnamefont{Wang}},
  \bibinfo{journal}{Nanoscale} \textbf{\bibinfo{volume}{4}},
  \bibinfo{pages}{5077} (\bibinfo{year}{2012}).

\bibitem[{\citenamefont{Huang et~al.}(2019)\citenamefont{Huang, Huang, Liu,
  Yang, Hu, Trochu, and Causse}}]{huang2019fourier}
\bibinfo{author}{\bibfnamefont{W.}~\bibnamefont{Huang}},
  \bibinfo{author}{\bibfnamefont{Q.}~\bibnamefont{Huang}},
  \bibinfo{author}{\bibfnamefont{Y.}~\bibnamefont{Liu}},
  \bibinfo{author}{\bibfnamefont{J.}~\bibnamefont{Yang}},
  \bibinfo{author}{\bibfnamefont{H.}~\bibnamefont{Hu}},
  \bibinfo{author}{\bibfnamefont{F.}~\bibnamefont{Trochu}}, \bibnamefont{and}
  \bibinfo{author}{\bibfnamefont{P.}~\bibnamefont{Causse}},
  \bibinfo{journal}{Comput. Method. Appl. M.} \textbf{\bibinfo{volume}{345}},
  \bibinfo{pages}{1114} (\bibinfo{year}{2019}).

\bibitem[{\citenamefont{Cerda}(2005)}]{cerda2005mechanics}
\bibinfo{author}{\bibfnamefont{E.}~\bibnamefont{Cerda}}, \bibinfo{journal}{J.
  Biomech.} \textbf{\bibinfo{volume}{38}}, \bibinfo{pages}{1598}
  (\bibinfo{year}{2005}).

\bibitem[{\citenamefont{David and Humphrey}(2004)}]{david2004redistribution}
\bibinfo{author}{\bibfnamefont{G.}~\bibnamefont{David}} \bibnamefont{and}
  \bibinfo{author}{\bibfnamefont{J.}~\bibnamefont{Humphrey}},
  \bibinfo{journal}{J. Biomech.} \textbf{\bibinfo{volume}{37}},
  \bibinfo{pages}{1197} (\bibinfo{year}{2004}).

\bibitem[{\citenamefont{Snyman and Wilke}(2005)}]{snyman2005practical}
\bibinfo{author}{\bibfnamefont{J.~A.} \bibnamefont{Snyman}} \bibnamefont{and}
  \bibinfo{author}{\bibfnamefont{D.~N.} \bibnamefont{Wilke}},
  \emph{\bibinfo{title}{Practical Mathematical Optimization}}
  (\bibinfo{publisher}{Springer}, \bibinfo{address}{New York},
  \bibinfo{year}{2005}).

\bibitem[{\citenamefont{Yao and Olvera de~la
  Cruz}(2016)}]{yao2016electrostatics}
\bibinfo{author}{\bibfnamefont{Z.}~\bibnamefont{Yao}} \bibnamefont{and}
  \bibinfo{author}{\bibfnamefont{M.}~\bibnamefont{Olvera de~la Cruz}},
  \bibinfo{journal}{Phys. Rev. Lett.} \textbf{\bibinfo{volume}{116}},
  \bibinfo{pages}{148101} (\bibinfo{year}{2016}).

\bibitem[{\citenamefont{Landau and Lifshitz}(1986)}]{Landau1986}
\bibinfo{author}{\bibfnamefont{L.~D.} \bibnamefont{Landau}} \bibnamefont{and}
  \bibinfo{author}{\bibfnamefont{E.~M.} \bibnamefont{Lifshitz}},
  \emph{\bibinfo{title}{Theory of Elasticity, 3rd edition}}
  (\bibinfo{publisher}{Butterworth-Heinemann}, \bibinfo{year}{1986}).

\bibitem[{\citenamefont{Steigmann}(1986)}]{steigmann1986proof}
\bibinfo{author}{\bibfnamefont{D.}~\bibnamefont{Steigmann}},
  \bibinfo{journal}{J. Appl. Mech.} \textbf{\bibinfo{volume}{53}},
  \bibinfo{pages}{955} (\bibinfo{year}{1986}).

\bibitem[{\citenamefont{Born}(1940)}]{born1940}
\bibinfo{author}{\bibfnamefont{M.}~\bibnamefont{Born}}, \bibinfo{journal}{Math.
  Proc. Cambridge Philos. Soc.} \textbf{\bibinfo{volume}{36(2)}},
  \bibinfo{pages}{160} (\bibinfo{year}{1940}).

\bibitem[{\citenamefont{Landau and Lifshitz}(1999)}]{Landau1999a}
\bibinfo{author}{\bibfnamefont{L.}~\bibnamefont{Landau}} \bibnamefont{and}
  \bibinfo{author}{\bibfnamefont{E.}~\bibnamefont{Lifshitz}},
  \emph{\bibinfo{title}{Statistical Physics}}
  (\bibinfo{publisher}{Butterworth-Heinemann}, \bibinfo{year}{1999}).

\bibitem[{\citenamefont{Miguel and Zapperi}(2003)}]{natmatcarmen2003}
\bibinfo{author}{\bibfnamefont{M.-C.} \bibnamefont{Miguel}} \bibnamefont{and}
  \bibinfo{author}{\bibfnamefont{S.}~\bibnamefont{Zapperi}},
  \bibinfo{journal}{Nat. Mater.} \textbf{\bibinfo{volume}{2}},
  \bibinfo{pages}{477} (\bibinfo{year}{2003}).

\bibitem[{\citenamefont{Miguel et~al.}(2011)\citenamefont{Miguel, Mughal, and
  Zapperi}}]{miguel2011laminar}
\bibinfo{author}{\bibfnamefont{M.-C.} \bibnamefont{Miguel}},
  \bibinfo{author}{\bibfnamefont{A.}~\bibnamefont{Mughal}}, \bibnamefont{and}
  \bibinfo{author}{\bibfnamefont{S.}~\bibnamefont{Zapperi}},
  \bibinfo{journal}{Phys. Rev. Lett.} \textbf{\bibinfo{volume}{106}},
  \bibinfo{pages}{245501} (\bibinfo{year}{2011}).

\end{thebibliography}

\end{document}